\documentclass[%
twocolumn,
superscriptaddress,
 amsmath,amssymb,
 aps,prb
]{revtex4-2}

\usepackage{graphicx}
\usepackage{dcolumn}
\usepackage{bm}
\usepackage{color}
\usepackage{ulem}
\usepackage{xspace}
\usepackage{multirow}
\usepackage{hyperref}
\usepackage{physics}
\usepackage{here}

\definecolor{green}{rgb}{0,0.6,0.1}

\newcommand{\QE}{{\textsc{Quantum ESPRESSO}}\xspace}

\begin{document}

\preprint{APS/123-QED}

\title{Efficient {\it ab initio} Migdal-Eliashberg calculation considering the retardation effect in phonon-mediated superconductors}
\author{Tianchun Wang}
\email{tiachun.wang@riken.jp}
\author{Takuya Nomoto}
\affiliation{
Department of Applied Physics, The University of Tokyo,7-3-1 Hongo, Bunkyo-ku, Tokyo 113-8656
}
\author{Yusuke Nomura}
\affiliation{ 
RIKEN Center for Emergent Matter Science, 2-1 Hirosawa, Wako, Saitama 351-0198, Japan 
}

\author{Hiroshi Shinaoka}
\affiliation{Department of Physics, Saitama University, Sakura, Saitama 338-8570, Japan}

\author{Junya Otsuki}
\affiliation{Research Institute for Interdisciplinary Science, Okayama University, Okayama 700-8530, Japan}

\author{Takashi Koretsune}
\affiliation{Department of Physics, Tohoku University, Miyagi 980-8578, Japan}

\author{Ryotaro Arita}
\affiliation{
Department of Applied Physics, The University of Tokyo,7-3-1 Hongo, Bunkyo-ku, Tokyo 113-8656
}
\affiliation{ 
RIKEN Center for Emergent Matter Science, 2-1 Hirosawa, Wako, Saitama 351-0198, Japan 
}


\date{\today}

\begin{abstract}
We formulate an efficient scheme to perform Migdal-Eliashberg calculation considering the retardation effect from first principles. While the conventional approach requires a huge number of Matsubara frequencies, we show that the intermediate representation of the Green's function [H. Shinaoka {\it et al.}, Phys. Rev. B 96, 035147 (2017)] dramatically reduces the numerical cost to solve the linearized gap equation. Without introducing any empirical parameter, we demonstrate that we can  successfully reproduce the experimental superconducting transition temperature of elemental Nb ($\sim 10$ K) very accurately. The present result indicates that our approach has a superior performance for many superconductors for which  $T_{\rm c}$ is lower than ${\mathcal O}(10)$ K.
\end{abstract}

\maketitle


\section{Introduction} 

{\it Ab initio} calculation of the superconducting transition temperature ($T_{\rm c}$) has been an intriguing challenge in computational condensed matter physics~\cite{Allen_1982,FLORESLIVAS2020}. Based on the experimental phonon spectrum of elemental Nb and the Migdal-Eliashberg theory, McMillan~\cite{McMillan_1968} and Allen and Dynes~\cite{Allen_1975} derived a formula to calculate $T_{\rm c}$ of phonon-mediated superconductors. While the McMillan-Allen-Dynes formula has been widely used to estimate $T_{\rm c}$ of a variety of superconductors~\cite{Allen_1982}, one crucial problem is that it contains a notorious empirical parameter, the pseudo Coulomb interaction parameter $\mu^{*}$~\cite{Morel_1962}. Thus McMillan-Allen-Dynes formula cannot be used for predicting $T_{\rm c}$ of unknown superconductors. 

On the other hand, it has been well known that the numerical cost for {\it ab initio} momentum-dependent Migdal-Eliashberg calculation is formidably expensive~\cite{Giustino_2008,GiustinoRMP2017}. Especially when $T_{\rm c}$ is relatively low, the retardation effect~\cite{Morel_1962} becomes more and more difficult to treat numerically~\cite{FLORESLIVAS2020}. While the typical energy scale of the dynamical structure of the electron-phonon interaction is just 10 meV, that of the screened Coulomb interaction is as large as the band width ($W\sim$10 eV). Thus in the scheme based on the Matsubara Green's function, we need to introduce an extremely large number of Matsubara frequencies (${N_{\rm M}}$) to describe the frequency dependence of the effective interaction between electrons accurately. In many cases, $N_{\rm M}$ should be  as large as $W/T$ to obtain a well-converged solution, where $T$ is the temperature. Therefore, {\it ab initio} Migdal-Eliashberg calculation requires considerably large memory and computation time, and has been performed only for hydride superconductors having $T_{\rm c}\sim 200$ K under high pressures~\cite{FLORESLIVAS2020,Sano2016,Errea2020}.

Besides the approach based on the Migdal-Eliashberg theory, there is another approach based on an extension of density functional theory, which is the so-called superconducting density functional theory (SCDFT)~\cite{Oliveira1988,Kreibich2001,SCDFT1,SCDFT2,Sanna_SCDFT2017}. In SCDFT, the gap equation consists of static quantities which do not depend on frequency, so that we can solve the gap equation in SCDFT much more efficiently than that in the Migdal-Eliashberg theory. Indeed, SCDFT calculations have been performed for many conventional superconductors~\cite{FLORESLIVAS2020,SCDFT1,SCDFT2,Sanna_SCDFT2017,kawamura2019,Akashi_2015,Jose_2016,Arita_2016}. 

In Ref.~\cite{SCDFT1}, the kernel of the SCDFT gap equation was constructed based on the Kohn-Sham perturbation theory. There, to describe the mass enhancement effect due to the electron-phonon coupling, the bare Green's function rather than the fully-dressed Green's function was employed. Therefore, the treatment of the mass enhancement effect in Ref.~\cite{SCDFT1} is not self-consistent. In fact, it is a highly non-trivial challenge to derive an exchange correlation functional considering the mass enhancement effect self-consistently. Thus the development of an efficient scheme to perform Migdal-Eliashberg calculation based on the fully-dressed Green's function is highly desired.

Recently, a method that can solve the long-standing problem of a large number of Matsubara frequencies has been developed~\cite{Otsuki_2020}. This method is based on a compact and efficient representation (which we call intermediate-representation (IR)) of the Green's functions proposed by two of the present authors and their collaborators~\cite{Shinaoka17,IRbasis_2018, IRbasis_2019, IRbasis_2020, Otsuki_2020}. The IR basis provides us not only a compact representation of the Green's function, but also enable us to perform efficient many-body calculations with the Green's functions. The number of the basis functions (the IR basis) required to store and reconstruct the Green's functions both in imaginary-time space and Matsubara-frequency space is much smaller than that of the conventional Legendre polynomials. There, it has been shown that the conventional uniform Matsubara-frequency grid can be replaced by a series of sparse sampling points
to describe the frequency dependence of the IR basis and hence Green's functions~\cite{IRbasis_2020}.
Using the sparse sampling method, we can reconstruct 
the Matsubara Green's function with only about 100 points on the frequency grid, and transform efficiently the imaginary-time Green's function to the Matsubara Green's function and vice versa. 

In this paper, we formulate a scheme to perform {\it ab initio} Migdal-Eliashberg calculation with the IR basis.  
We find that we can solve the anisotropic (momentum-dependent) gap equation very efficiently. We show the results for two different superconductors: One is elemental Nb with $T_{\rm c}\sim$ 10 K, and the other is LaH$_{10}$ under 250 GPa with $T_{\rm c}\sim$ 200 K~\cite{Errea2020}. With these benchmark calculations, we demonstrate that our new approach has a superior performance especially when $T_{\rm c}$ is lower than ${\mathcal O}(10)$ K.  

\section{Method}
\label{sec_methods}
\subsection{Eliashberg equation}
\label{subsec_Eliashberg}
In the framework of the Migdal-Eliashberg theory~\cite{Migdal_1958,FLORESLIVAS2020,Eliashberg_1960, Allen_1982}, we calculate the superconducting $T_{\rm c}$ by solving the gap equation,
\begin{align}
\label{eqn:Eliashberg}
\Delta_m(\bm{k},i{\omega}_n)&= -\frac{T}{N_{\bm{k}}}\sum_{m^{\prime}}\sum_{\bm{k}^{\prime},i{\omega}_{n^{\prime}}}\mathcal{K}_{mm^{\prime}}(\bm{k}-\bm{k}^{\prime},i{\omega}_n-i{\omega}_{n^{\prime}}) \nonumber
\\
&\hspace{3cm}{\times}F_{m^{\prime}}(\bm{k}^{\prime},i{\omega}_{n^{\prime}}),
\end{align}
where $\Delta_m$ is the superconducting gap function, $\mathcal{K}_{mm^\prime}$ a pairing-interaction kernel, and $F_{m^\prime}$ the  anomalous Green's function, which are functions of the electron momenta $\bm{k},{\bm k'}$, Matsubara frequencies $\omega_n$, $\omega_{n^\prime}$, and band indices $m,m^\prime$. $N_{\bm{k}}$ denotes the total number of ${\bm k}$-points. $T_{\rm c}$ is the highest temperature $T$ at which $\Delta_m$ is finite. 

In the following calculations, we take the linear approximation that the second order products of the anomalous quantities will be ignored, then the anomalous Green's function can be written as,
\begin{equation}
\label{eqn:anomalous}
F_{m}(\bm{k},i{\omega}_{n})=|G_{m}(\bm{k},i{\omega}_{n})|^{2}{\Delta}_{m}(\bm{k},i{\omega}_{n}),
\end{equation}
where $G_{m}(\bm{k},i{\omega}_{n})$ is the electron Green's function. The kernel $\mathcal{K}_{mm^{\prime}}$ consists of the contributions from the attractive interaction due to electron-phonon coupling and the repulsive screened Coulomb interaction,
\begin{equation}
\label{eqn:kernel}
\mathcal{K}_{mm^{\prime}}=\mathcal{K}^{\rm el\mathchar`-ph}_{mm^{\prime}}+\mathcal{K}^{\rm C}_{mm^{\prime}}. 
\end{equation}

Let us first focus on the first term and leave the treatment of the second term in Sec.~\ref{subsec_Coulomb}. Considering the electron-phonon interaction as a scattering process of electrons from a momentum $\bm{k}$ to $\bm{k}-\bm{q}$ mediated by phonons with a momentum $\bm{q}$, we can write
$\mathcal{K}^{\rm el\mathchar`-ph}_{mm^{\prime}}$ as,
\begin{align}
\mathcal{K}^{\rm el\mathchar`-ph}_{mm^{\prime}}(\bm{q},i{\omega}_{\nu})=\frac{1}{N_k}\sum_{\lambda,\bm{k}}|\textsl{g}_{\lambda}^{m\bm{k},m^{\prime}\bm{k}-\bm{q}}|^2D_{\lambda}(\bm{q},i{\omega}_{\nu}),
\end{align}
where $\lambda$ and ${\omega}_{\nu}$ are phonon's mode index and Matsubara frequency of bosons, respectively. We assume that the electron-phonon interaction matrix element $\textsl{g}_{\lambda}^{m\bm{k},m^{\prime}\bm{k}-\bm{q}}$ does not depend on $\bm{k}$ significantly, and take an average over $\bm{k}$ on the Fermi level.
The phonon Green's function
$D_{\lambda}(\bm{q},i{\omega}_{\nu})$ 
is given as,
\begin{equation}
D_{\lambda}(\bm{q},i{\omega}_{\nu})= -\frac{2{\omega}_{\bm{q}{\lambda}}}{{\omega}_{\nu}^2+{\omega}_{\bm{q}{\lambda}}^2},
\end{equation}
where ${\omega}_{\bm{q}{\lambda}}$ is an energy dispersion of phonons.
In the present study, we calculate ${\omega}_{\bm{q}{\lambda}}$ and $\textsl{g}_{\lambda}^{m\bm{k},m^{\prime}\bm{k}-\bm{q}}$ by density functional perturbation theory (DFPT)~\cite{Baroni_2001}. 

Before the electron Green's function $G_m(\bm{k},i{\omega}_n)$ enters Eqs.~(\ref{eqn:Eliashberg}) and (\ref{eqn:anomalous}), we consider the self-energy due to the electron-phonon interaction,
\begin{align}
\label{eqn:Self-energy}
\Sigma_m(\bm{k},i{\omega}_n)= -\frac{T}{N_{\bm{k}}}\sum_{m^{\prime}}\sum_{\bm{k}^{\prime},i{\omega}_{n^{\prime}}}\mathcal{K}^{\rm el\mathchar`-ph}_{mm^{\prime}}(\bm{k}-\bm{k}^{\prime},i{\omega}_n-i{\omega}_{n^{\prime}}) \nonumber
\\
&\hspace{-3.8cm}{\times}G_{m^{\prime}}(\bm{k}^{\prime},i{\omega}_{n^{\prime}}).
\end{align}
By solving the Dyson equation self-consistently, we obtain the dressed electron Green's function,
\begin{equation}
\label{eqn:Dyson}
G_m(\bm{k},i{\omega}_n)= \frac{1}{i{\omega}_{n}-{\varepsilon}_{{m}\bm{k}}-{\Sigma}_m(\bm{k},i{\omega}_n)},
\end{equation}
where ${\varepsilon}_{{m}\bm{k}}$ is the bare energy dispersion of electrons. If we have $ab$ $initio$  results for ${\varepsilon}_{{m}\bm{k}}$, ${\omega}_{\bm{q}{\lambda}}$, $\textsl{g}_{\lambda}^{m\bm{k},m^{\prime}\bm{k}-\bm{q}}$, and $\mathcal{K}^{\rm C}_{mm^{\prime}}$, we can solve Eq.~(\ref{eqn:Eliashberg}) and calculate $T_{\rm c}$ from first principles. It should be noted that Eq.~(\ref{eqn:Eliashberg}) with the approximation \eqref{eqn:anomalous} becomes an eigenvalue problem,
\begin{align}
\label{eqn:Gapeq}
{\tilde\lambda}\Delta_m(\bm{k},i{\omega}_n)&= -\frac{T}{N_{\bm{k}}}\sum_{m^{\prime}}\sum_{\bm{k}^{\prime},i{\omega}_{n^{\prime}}}\mathcal{K}_{mm^{\prime}}(\bm{k}-\bm{k}^{\prime},i{\omega}_n-i{\omega}_{n^{\prime}}) \nonumber
\\
&{\times}|G_{m^\prime}(\bm{k}^\prime,i{\omega}_{n^\prime})|^{2}{\Delta}_{m^\prime}(\bm{k}^\prime,i{\omega}_{n^\prime}).
\end{align}
Here we introduce a parameter ${\tilde\lambda}$ as an eigenvalue. Using the power method, we calculate ${\tilde \lambda}$ for different temperatures. The maximum eigenvalue ${\tilde \lambda}_{\rm max}$ reaches unity when  $T=T_{\rm c}$.

\subsection{Screened Coulomb interaction}
\label{subsec_Coulomb}
In the present study, following the SCDFT calculation~\cite{SCDFT2}, we employ the static approximation for the screened Coulomb interaction which successfully reproduces the experimental $T_{\rm c}$ of elemental Nb. It should be noted that while the plasmon effect enhances $T_{\rm c}$~\cite{Akashi_2013,Akashi_2014}, spin fluctuations suppress $T_{\rm c}$~\cite{Essenberger_SpinFluctuationsTheory_PRB2014,Essenberger_FeSe_PRB2016,kawamura2019}. Thus SCDFT calculation considering these effects gives similar $T_{\rm c}$ to that of the static approximation~\cite{kawamura2019}.

Based on the results of DFT calculations, the polarizability function in the random phase approximation (RPA) can be written as~\cite{aulbur2000quasiparticle}
\begin{align}
\chi_{\bm{G}\bm{G^{\prime}}}(\bm{q},i\omega_\nu) = \frac{2}{N_{\bm{k}}}\sum_{m,m^{\prime}}\sum_{\bm{k}}M^{\bm{G}}_{{m}{\bm{k}+\bm{q}},m^{\prime}\bm{k}}M^{\bm{G^{\prime}}{*}}_{{m}{\bm{k}+\bm{q}},m^{\prime}\bm{k}}
\nonumber
\\
&\hspace{-3.8cm}{\times}X_{{m}{\bm{k}+\bm{q}},m^{\prime}{\bm{k}}}(i\omega_\nu),
\end{align}
with interstate scattering matrix
\begin{align}
M^{\bm{G}}_{{m}{\bm{k}+\bm{q}},m^{\prime}\bm{k}}=\mel{\psi_{m\bm{k}+\bm{q}}}{e^{i(\bm{q}+\bm{G})\cdot\bm{r}}}{\psi_{m^{\prime}{\bm{k}}}},
\end{align}
as well as
\begin{align}
X_{{m}{\bm{k}+\bm{q}},m^{\prime}{\bm{k}}}(i\omega_\nu)=\frac{f_{{m}{\bm{k}+\bm{q}}}-f_{m^{\prime}{\bm{k}}}}{i\omega_\nu+(\varepsilon_{{m}{\bm{k}+\bm{q}}}-\varepsilon_{m^{\prime}{\bm{k}}})},
\end{align}
where $\psi_{m\bm{k}+\bm{q}}$ and $\psi_{m^{\prime}{\bm{k}}}$ represent the Kohn-Sham wave function, 
$f_{{m}{\bm{k}+\bm{q}}}$ and $f_{m^{\prime}{\bm{k}}}$ are the corresponding Fermi distribution function, and $\bm{G}$ and $\bm{G}^{\prime}$ are reciprocal lattice vectors, respectively.
Then the RPA dielectirc function is 
\begin{align}
\epsilon_{\bm{G}\bm{G^{\prime}}}(\bm{q},i\omega_\nu) = \delta_{\bm{G}\bm{G^{\prime}}}-\frac{4\pi}{\Omega}\frac{1}{|\bm{q}+\bm{G}|}\chi_{\bm{G}\bm{G^{\prime}}}(\bm{q},i\omega_\nu)\frac{1}{|\bm{q}+\bm{G^{\prime}}|}.
\end{align}
With the Fourier transformation of the dielectric function, combined with the bare Coulomb interaction, we can write the screened Coulomb interaction
\begin{align}
w(\bm{r},\bm{r}^{\prime},i\omega_\nu)=\int_V d\bm{r}^{{\prime}{\prime}}\frac{\epsilon^{-1}(\bm{r},\bm{r}^{{\prime}{\prime}},i\omega_\nu)}{|\bm{r}^{\prime}-\bm{r}^{{\prime}{\prime}}|},
\end{align}
as well as its scattering matrix elements between two Kohn-Sham electrons as
\begin{align}
\label{eqn:W_RPA}
W_{m\bm{k},m^{\prime}\bm{k^{\prime}}}^{\rm RPA}(i\omega_\nu) = \int_V d\bm{r} \int_V d\bm{r^{\prime}}\psi_{m\bm{k}}^*(\bm{r})\psi_{m^{\prime}\bm{k^{\prime}}}(\bm{r})w(\bm{r},\bm{r^{\prime}},i\omega_\nu) \nonumber
\\
&\hspace{-4.2cm}{\times}\psi_{m^{\prime}\bm{k^{\prime}}}^*(\bm{r^{\prime}})\psi_{m\bm{k}}(\bm{r^{\prime}}).
\end{align}
Therefore we can write the Coulomb kernel in Eq.~(\ref{eqn:kernel}) as the static mode of RPA screened Coulomb interaction:
\begin{align}
\label{eqn:W_static}
\mathcal{K}^{\rm C}_{mm^{\prime}}(\bm{q},i{\omega}_{\nu})=W^{\rm RPA}_{mm^{\prime}}(\bm{q},i\omega_{\nu}=0),
\end{align}
where $\bm{q}=\bm{k}-\bm{k^{\prime}}$, and we have neglected the plasmon effect, namely, the Matsubara frequency dependence of the screen Coulomb interactions. Taking average of $W^{\rm RPA}_{m\bm{k},m^{\prime}\bm{k^{\prime}}}(i{\omega}_{\nu}=0)$ over the Fermi surface, we will get a parameter ${\mu}_{\rm C}$ to estimate the effect of screened Coulomb interaction of the system as
\begin{align}
\label{eqn:mu}
{\mu}_{\rm C}N(0)=\sum_{m\bm{k},m^{\prime}\bm{k^{\prime}}}W^{\rm RPA}_{m\bm{k},m^{\prime}\bm{k^{\prime}}}(i{\omega}_{\nu}=0)\delta({\varepsilon}_{{m}\bm{k}})\delta({\varepsilon}_{m^{\prime}\bm{k^{\prime}}}),
\end{align}
where $N(0)$ is the total density of states at the Fermi level.

\subsection{Fourier transformation with the IR basis}
\label{subsec_FFT}
When we solve Eqs.~(\ref{eqn:Self-energy}) and (\ref{eqn:Gapeq}), we have to calculate the convolution of $\mathcal{K}^{\rm el\mathchar`-ph}_{mm^{\prime}}$ and $G_{m^\prime}$ and
that of $\mathcal{K}_{m m^{\prime}}$ and $|G_{m^\prime}|^2\Delta_{m^\prime}$, respectively.
In general, we can write the convolution of two functions $f$ and $g$ on the discrete imaginary frequency grid $\{i{\omega}_{n}\}$ as
\begin{align}
\label{eqn:convolution}
\sum_{i{\omega}_{n^{\prime}}}f(i{\omega}_{n}-i{\omega}_{n^{\prime}})g(i{\omega}_{n^{\prime}}) = \mathcal{F}^{-1}[\mathcal{F}(f)*\mathcal{F}(g)],
\end{align}
where $\mathcal{F}$ and $\mathcal{F}^{-1}$ are the Fourier and inverse Fourier transform between the imaginary frequency space $\{i{\omega}_{n}\}$ and imaginary time space $\{\tau_m\}$, and star $*$ denotes the dot product of two arrays. The convolution on the discrete $\bm{k}$ mesh can be calculated similarly.
In the conventional calculation of $T_{\rm c}$~\cite{Sano2016,Errea2020}, we calculate Eq.~(\ref{eqn:convolution}) using the fast Fourier transformation (FFT) 
~\cite{FFTW_2005}.

However, for systems with relatively low $T_{\rm c} \lesssim 10$ K, the conventional FFT method will always encounter a problem of expensive computational cost. This is because the uniform Matsubara frequency grid of $\{i\omega_n\}$ becomes denser and denser at low temperature, while the cutoff frequency is always as high as the band width $W \gtrsim 10$ eV.
Thus we need to introduce a huge number of Matsubara frequencies to perform a calculation for $T\sim 10$ K. 

To overcome this problem, in the present study, we introduce an alternative route combining the FFT and the IR basis~\cite{Shinaoka17,IRbasis_2018,IRbasis_2019,IRbasis_2020, Otsuki_2020}. 
With the pre-computed IR basis functions $\{U_l^{\alpha}\}$~\cite{IRbasis_2019}, we have a compact and efficient representation of the Matsubara Green's function:
\begin{align}
\label{eqn:IR-basis1}
G^{\alpha}(i{\omega}_{n})=\sum_{l=0}^{l_{\rm max}}{G_l^{\alpha}U_l^{\alpha}(i{\omega}_{n})},
\end{align}
\begin{align}
\label{eqn:IR-basis2}
G^{\alpha}({\tau}_{m})=\sum_{l=0}^{l_{\rm max}}{G_l^{\alpha}U_l^{\alpha}({\tau}_{m})},
\end{align}
where $\alpha$ = F,B denotes the fermionic and bosonic Green's functions, respectively. The expansion of the Green's functions using the IR basis depends on two dimensionless parameters $\Lambda$ and $l_{\rm max}$, where $\Lambda=\beta\omega_{\rm max}$, $\beta=1/T$ is an inverse temperature, and $\omega_{\rm max}$ is a cutoff frequency of the spectral function. The value of $\Lambda$ controls the truncation errors due to the frequency window.
The number of basis functions $l_{\rm max}$ grows only logarithmically with respect to $\Lambda$~\cite{IRbasis_2018} (e.g., typical value of $l_{\rm max}$ for $\Lambda=10^5$ is 136, and $l_{\rm max}$ for $\Lambda=10^7$ is 201).
In the present study for Nb and LaH$_{10}$, we need no more than 200 IR basis functions.

We solve from Eqs.~(\ref{eqn:Self-energy}) to (\ref{eqn:Gapeq}) by means of the sparse sampling method based on the IR basis~\cite{IRbasis_2020}.
In this method, one takes sampling points in imaginary time $\{\bar{\tau}^\alpha_k\}$ ($k=0, 1, \cdots$)
according to the distribution of $l_{\rm max}$ roots of the highest order basis function $U_{l_{\rm max}}^{\alpha}({\tau})$.
Similarly, one takes Matsubara frequency sampling points $\{i\bar{\omega}^\alpha_k\}$ ($k=0, 1, \cdots$) according to the distribution of the sign changes of $U_{l_{\rm max}}^{\alpha}(i\omega_n)$.
This procedure always generates $l_{\rm max}$ or $l_{\rm max}+1$ sampling points, whose distribution depends on the statistics $\alpha$ and $\Lambda$ by construction.
The sampling points are sparsely and non-uniformly distributed, covering from low to high frequency regions more efficiently than uniform grids.

When $i\omega_n$ is replaced by $i\bar{\omega}^\alpha_k$ in Eq.~(\ref{eqn:IR-basis1}), $U_l^{\alpha}(\{i{\bar{\omega}}_k^\alpha\})$ can be regarded as a matrix element with the dimension index $l$ and sampling point index $k$.
Thus, 
one can evaluate the expansion coefficients $G^\alpha_l$ from $G^\alpha(i\omega_n)$ given on the sampling points by a least-squares fitting procedure with a precomputed (pseudo) inverse of the fitting matrix.
This procedure is numerically stable because the sampling points are chosen so as to minimize the condition number of the fitting matrix.
The inverse transform from the right-hand side to the left-hand side is a simple matrix multiplication.

These two transforms together with their counterparts for the $\tau$ sampling enable efficient transforms  between the imaginary-time and Matsubara-frequency space via the IR basis.
They are much more efficient than the conventional FFT method.
For example, in Sec.~\ref{subsec_convergence}, we will demonstrate that in the calculation of Nb with $T_{\rm c}\sim 10$ K, the size of the fitting matrix is around 150$\times$150 with $\Lambda=10^5$, while the FFT requires at least 4000 Matsubara frequencies to give comparable results.
We refer the reader to Ref.~\cite{IRbasis_2020} for more technical details on the sparse sampling method.

\subsection{Calculation conditions}
\label{subsec_cond}
In this paper, we calculate $T_{\rm c}$ of elemental Nb and LaH$_{10}$. Elemental Nb has the body-centered-cubic lattice. The lattice parameter is optimized as $a = 3.31$\AA\ where the experimental value is $a=3.30$\AA. Following  Ref.~\cite{Errea2020}, we take the crystal structure of the $Fm{\bar 3}m$ face-centered-cubic phase of LaH$_{10}$ at 250 GPa with lattice constant $a = 4.84$\AA . For the DFT calculation, we use \QE code~\cite{QE-2017} with the exchange correlation functional proposed by Perdew, Burke, and Ernzerhof~\cite{Perdew_1996}. We use a projector-augmented wave (PAW)~\cite{PAW_1994} pseudopotential for Nb and  ultrasoft pseudopotentials~\cite{USPP_1990} for La and H atoms. All these pesudopotentials are provided in PSLibrary~\cite{PSLibrary_1.0.0}.
The cutoff energy for the plane wave expansion of the wave functions is set to be 70 Ry for Nb and 50 Ry for LaH$_{10}$. The cutoff for the charge density are 280 Ry for Nb and 500 Ry for LaH$_{10}$. For the DFPT calculation, we use the package in 
\QE~\cite{QE-2017}. For Nb, we take a
20$\times$20$\times$20 and  18$\times$18$\times$18 $\bm{k}$-mesh
for a 10$\times$10$\times$10 
and 9$\times$9$\times$9 $\bm{q}$-mesh, respectively. For LaH$_{10}$, we use a 12$\times$12$\times$12 $\bm{k}$-mesh and a 6$\times$6$\times$6 $\bm{q}$-mesh.

In the calculation of Eqs.~(\ref{eqn:Self-energy}) and (\ref{eqn:Gapeq}), we use the conventional FFT to take a convolution on the $\bm{k}$ mesh, and use the IR-basis to take a convolution on the Matsubara frequency grid. 
For Nb, we use the number of $\bm{k}$ points ranging from 36$\times$36$\times$36 to 100$\times$100$\times$100 for sampling in the first Brillouin zone to check the convergence. We use a 36$\times$36$\times$36 $\bm{k}$ mesh for LaH$_{10}$. 

For the calculation of the screened Coulomb interaction for Nb, we use a 18$\times$18$\times$18 and 20$\times$20$\times$20 $\bm{k}$-mesh for a 9$\times$9$\times$9 and 10$\times$10$\times$10 $\bm{q}$-mesh, respectively. 
20 unoccupied bands are used for Nb. For LaH$_{10}$, we use a 12$\times$12$\times$12 $\bm{k}$-mesh, a 6$\times$6$\times$6 $\bm{q}$-mesh and 30 unoccupied bands. The cutoff energy for the dielectric function is set to be 70 Ry for Nb and 50 Ry for LaH$_{10}$.
The resulting Coulomb parameter in Eqs.~(\ref{eqn:mu}) is 0.24  for LaH$_{10}$, and 0.43 for Nb. In the following calculations, we use $W^{\rm RPA}_{mm^{\prime}}(\bm{q},i{\omega}_{\nu}=0)$ as the screened Coulomb kernel.
For computing IR basis functions and the sampling points,
we use the irbasis library~\cite{IRbasis_2019}. 

\section{Result and Discussion}
\label{sec_results}
\subsection{Convergence along Matsubara frequencies} 
\label{subsec_convergence}
\begin{figure}[tb]
\vspace{0cm}
\begin{center}
\includegraphics[width=0.48\textwidth]{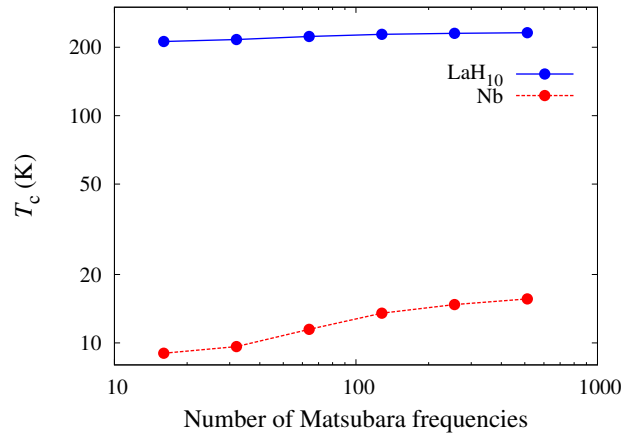}
\caption{
$N_{\rm M}$ (number of Matsubara frequencies) dependence of $T_{\rm c}$ for Nb and LaH$_{10}$ at 250 GPa. Results are shown in a logarithmic scale. $T_{\rm c}$ for Nb is calculated using a 9$\times$9$\times$9 $\bm{q}$ mesh and a 36$\times$36$\times$36 $\bm{k}$ mesh. Data points are connected by lines.
}
\label{tc_vs_matsubara_LaH10}
\end{center}
\end{figure}

\begin{figure*}[tb]
\centering
    \begin{minipage}[b]{0.48\textwidth}
      \centering 
      \includegraphics[width=\linewidth]{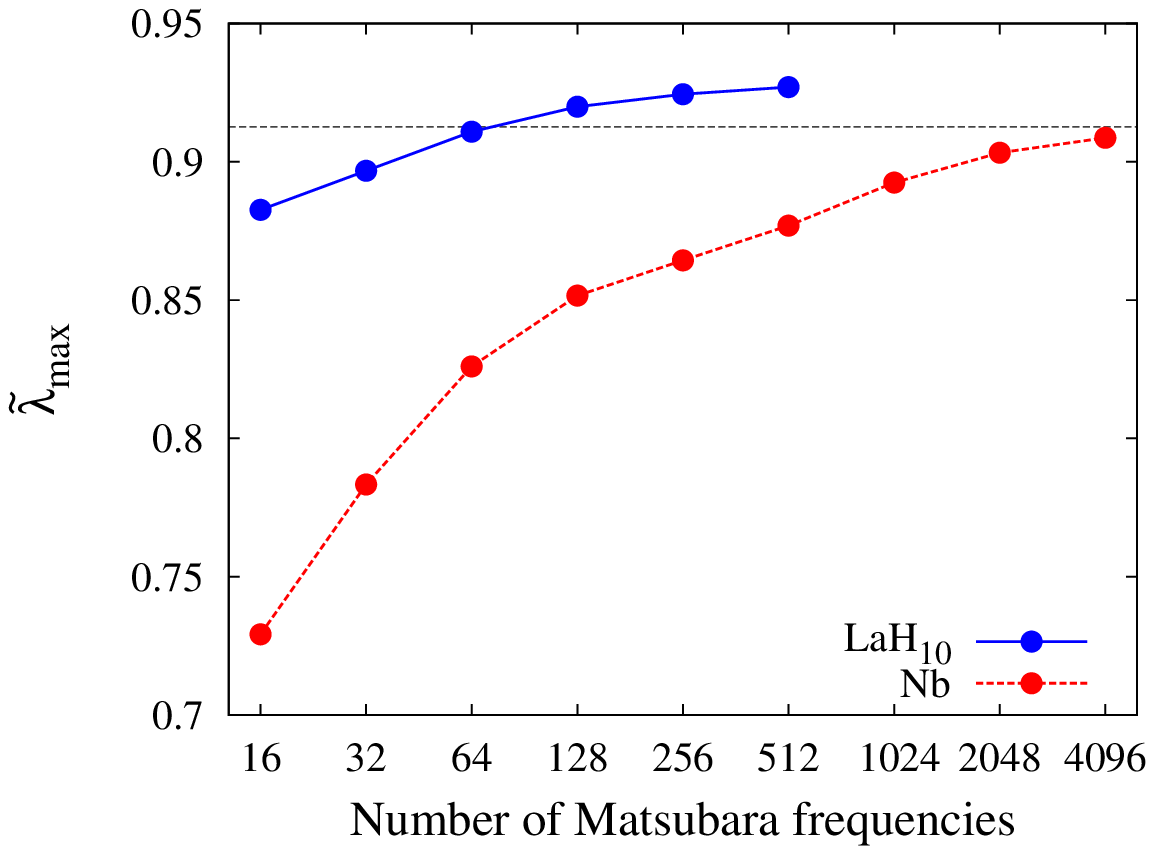}
    \end{minipage}
        \begin{minipage}[b]{0.48\textwidth}
      \centering 
      \includegraphics[width=\linewidth]{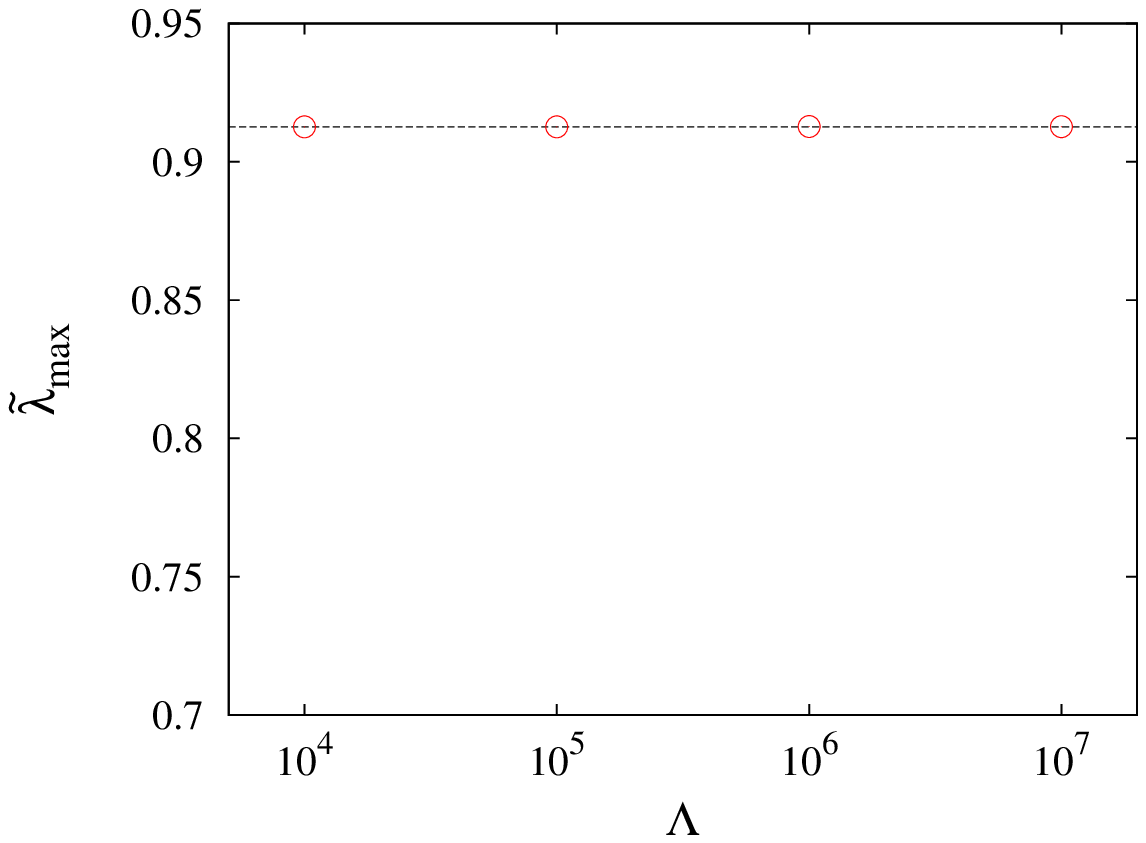}
    \end{minipage}
\caption{Left panel: Eigenvalue ${\tilde \lambda}_{\rm max}$ vs number of Matsubara frequencies ($N_{\rm M}$) for Nb and LaH$_{10}$ at 250 GPa. Based on the conventional FFT method, eigenvalue is calculated with $T$ fixed at 19.7 K for Nb, and 271.6 K for LaH$_{10}$. Eigenvalue reaches numerical convergence with $N_{\rm M}=$ 512 for LaH$_{10}$, while it requires $N_{\rm M}=$ 4096 for Nb. Data points are connected by lines. Right panel: Eigenvalue ${\tilde \lambda}_{\rm max}$ for Nb vs dimensionless parameter $\Lambda$ in the IR basis method. Temperature $T$ is fixed at 19.7 K. The horizontal dashed lines in both panels indicate that the IR basis method gives ${\tilde \lambda}_{\rm max} = 0.913$ for Nb, which is consistent with the converged value of ${\tilde \lambda}_{\rm max}$ for $N_{\rm M}\rightarrow\infty$. Calculations for Nb in both panels are calculated using a 9$\times$9$\times$9 $\bm{q}$ mesh and a 36$\times$36$\times$36 $\bm{k}$ mesh.}
\label{compare}
\end{figure*}

To solve the Eliashberg equation based on the Matsubara Green's functions, a large $N_{\rm M}$ has to be employed in the calculation, which causes the numerical difficulty due to the expensive memory and computational time. Since the Matsubara frequencies are proportional to $T$, the required number of $N_{\rm M}$ linearly increases as decreasing $T$. Thus, it is extremely difficult to solve the equation in a system with low $T_{\rm c}$. In this section, we will demonstrate this problem by comparing calculation of Nb and LaH$_{10}$ at 250 GPa, one with $T_{\rm c}$ about 10 K, and the other one with a high $T_{\rm c}$ around 230 K.

Figure \ref{tc_vs_matsubara_LaH10} shows the numerical convergence of $T_{\rm c}$ for Nb as well as LaH$_{10}$, calculated with different numbers of $N_{\rm M}$. 
Since LaH$_{10}$ has a high $T_{\rm c}\sim230$ K, 
the whole range of energy bands are covered with only several hundred Matsubara frequencies, therefore the result of $T_{\rm c}$ reaches convergence. However, with the same number of Matsubara frequencies, we cannot get a converged result of $T_{\rm c}$ for Nb, because $T_{\rm c}$ for Nb is much lower.

In Fig.~\ref{compare}, we compare the convergence of ${\tilde \lambda}_{\rm max}$ for Nb and that for LaH$_{10}$ at 250 GPa (left panel), and also shows the results for Nb based on the IR basis method (right panel). For LaH$_{10}$, only a few hundred Matsubara frequencies are enough to obtain the converged value because $T$ is sufficiently high (271.6 K). On the other hand, for Nb, we need at least 4096 Matsubara frequencies to reach convergence, which is due to the low temperature used in the calculation (19.7 K). However, as is seen in the right panel of Fig.~\ref{compare}, ${\tilde\lambda}_{\rm max}$ of the IR basis method reaches convergence at $\Lambda=10^{4}$, which only requires $l_{\rm max}=103$ 
basis functions. Thus, it is obvious that the IR basis method performs better in Nb. Note that the IR basis method gives the same ${\tilde\lambda}_{\rm max}$ as the conventional method in the limit of the large $N_{\rm M}$. In addition, comparing the computational time for a single calculation of convolution, the IR basis method with $\Lambda=10^{5}$ performs 20 times faster than the conventional FFT method with 4096 Matsubara frequencies. In the following calculations, we set $\Lambda=10^{5}$, where $l_{\rm max}=136$ and the number of sampling Matsubara frequency points for fermions is 138.

\subsection{Critical temperature and gap function} 
\label{sec_tc_and_gap}

\begin{figure}[tb]
\vspace{0cm}
\begin{center}
\includegraphics[width=0.48\textwidth]{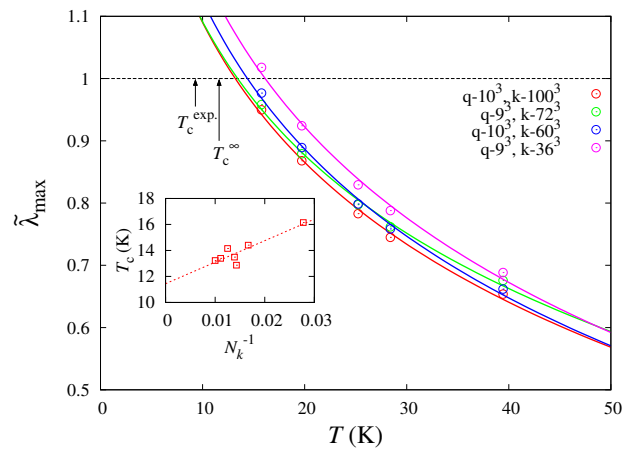}
\caption{
Eigenvalue ${\tilde \lambda}_{\rm max}$ vs temperature $T$ for Nb. Circles with different colors are results of different meshes with different number of sampling $\bm{k}$ points and $\bm{q}$ points. Solid lines with corresponding colors are fitting results using a fitting function ${\tilde \lambda}_{\rm max}$ = $A$+$B$log$T$. The crossing points of the fitting lines and horizontal dashed line ${\tilde \lambda}_{\rm max} = 1$ are $T_{\rm c}$. Numerical results of $T_{\rm c}$ for a 100$\times$100$\times$100 $\bm{k}$ mesh is 13.2 K. The inset shows $N_{k}^{-1}$ dependence of $T_{\rm c}$, as well as a linear fitting function as $T_{\rm c}$ = $T_{\rm c}^{\infty}$+a$N_{k}^{-1}$, where ${N_k}$ is the number of $\bm{k}$ points in one dimension. The extrapolation result of $T_{\rm c}$ with ${N_k}\to\infty$ is $T_{\rm c}^{\infty}= 11.4$ K. The experimental result of $T_{\rm c}$ for Nb is $T_{\rm c}^{\rm exp.}= 9.3$ K~\cite{Aschcroftsolid}.
}
\label{Tc}
\end{center}
\end{figure}

\begin{figure}[tb]
\vspace{0cm}
\begin{center}
\includegraphics[width=0.48\textwidth]{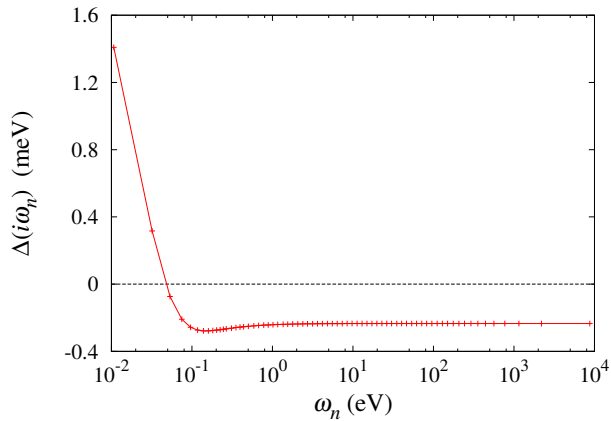}
\caption{
Normalized eigenfunction $\Delta$ in the Eliashberg equation for Nb as a function of Matsubara frequency $\omega_n$. Frequency sampling points are sparsely distributed along the frequency axis. Temperature is set to 39.4 K, and we use a 100$\times$100$\times$100 $\bm{k}$ mesh. The wave-number is fixed at $\Gamma$ point (0, 0, 0) in the Brillouin zone, and we choose the band just above the Fermi level. Data points are connected by lines.
}
\label{gap_Nb}
\end{center}
\end{figure}

Besides the convergence test for $N_{\rm M}$, we should also consider the convergence on a discrete $\bm{k}$ mesh. Numerical results of different $\bm{q}$ meshes and $\bm{k}$ meshes is shown in Fig.~\ref{Tc}. 
With the eigenvalue ${\tilde \lambda}$ in Eqs.~(\ref{eqn:Eliashberg}), we can solve the equation at different temperature. Then we can get a numerical results of $T_{\rm c}$, as shown in Fig.~\ref{Tc}. All the results of $T_{\rm c}$ with different number of sampling $\bm{k}$ points are shown in the inset of Fig.~\ref{Tc}. 
For the calculation with a 36$\times$36$\times$36 and 72$\times$72$\times$72 $\bm{k}$ mesh,
a 9$\times$9$\times$9 $\bm{q}$-mesh are used to calculate the screened Coulomb interaction. In the calculation on the other $\bm{k}$ meshes, the screened Coulomb interaction is calculated using 
a 10$\times$10$\times$10 $\bm{q}$-mesh. Linear interpolation is employed 
to use the screened Coulomb interaction data on the coarse $\bm{q}$-mesh 
in the Eliashberg calculation on the dense $\bm{k}$-mesh. $T_{\rm c}$ for 100$\times$100$\times$100 $\bm{k}$ mesh is 13.2 K. The deviation between the results with different $\bm{k}$-mesh calculations increases with lowering the temperature
since the discrete $\bm{k}$-mesh approximation becomes less accurate. We can expect that the numerical result will become closer to the experimental value with a much denser $\bm{k}$ mesh. A linear extrapolation of the results to the infinite number of sampling $\bm{k}$ points gives $T_{\rm c}^{\infty}= 11.4$ K, which successfully reproduces the experimental result of $T_{\rm c}^{\rm exp.}= 9.3$ K~\cite{Aschcroftsolid}. 

Although we have neglected the dynamical structure of the screened Coulomb kernel, our numerical result turns out to have a good agreement with the experimental $T_{\rm c}$, which is because we have neglected both the effects of plasmon and spin fluctuations. Since the effects of plasmon increase $T_{\rm c}$ and spin fluctuations decrease $T_{\rm c}$, these two effects on $T_{\rm c}$ will counteract with each other eventually.
A similar cancellation is also seen in the SCDFT; 
calculations using static~\cite{SCDFT2} and dynamical~\cite{kawamura2019} Coulomb kernel give comparable $T_{\rm c}$. 
We note that the value of $T_{\rm c}$ in Ref.~\onlinecite{SCDFT2} is slightly lower than that of the present study. 
This might be ascribed to the fact that the mass enhancement in the SCDFT is not calculated self-consistently. 
As is shown in Ref.~\cite{Sano2016}, within the Migdal-Eliashberg theory, the mass enhancement effect is overestimated in the one-shot calculation, and $T_{\rm c}$ tends to be lower than that of the self-consistent calculation.

In Fig.~\ref{gap_Nb}, we plot the normalized eigenfunction $\Delta$ of the Eliashberg equation~(\ref{eqn:Gapeq}) at $\bm{k}$ close to the Fermi level 
as a function of the Matsubara frequency. $\Delta(i\omega_n)$ changes rapidly in the range of $10^{-2}$ to $10^{-1}$ eV, which is a typical energy scale of the Debye frequency. This means that the behaviour of $\Delta(i\omega_n)$ in this scale is dominated by the electron-phonon interaction. $\Delta(i\omega_n)$ becomes negative for $\omega_n \gtrsim 10^{-1}$ eV due to the retardation effect, which has not been obtained in previous Migdal-Eliashberg calculations for low $T_{\rm c}$ superconductors.
It should be noted that the sampling frequency points in Fig.~\ref{gap_Nb}
are sparsely sampled along the imaginary frequency axis. 
This confirms our discussions in Sec.~\ref{subsec_FFT} that our scheme based on the IR basis can easily reach high energy region $\sim$10 eV
without introducing a huge number of Matsubara frequencies. 

\section{Conclusion} 
We have formulated a fully $ab$ $initio$ scheme to perform calculations on superconducting transition temperature $T_{\rm c}$, combining with the recently proposed intermediate-representation basis of the Green's function. With the consideration of the fully-dressed Green's function, our numerical result successfully reproduced the experimental result, without considerably large memory and computation time cost, which is always troublesome in the conventional approach. It provides an efficient and promising approach to calculate and predict properties of superconducting systems at $T\lesssim 10$K.

\begin{acknowledgments}
This work was supported by a Grant-in-Aid for Scientific Research (JP19H05825,  JP19K14654, JP18K03442, JP18H01158, JP16K17735, JP17K14336, JP20K14423 and JP16H06345) from Ministry of Education, Culture, Sports, Science and Technology.
\end{acknowledgments}

\appendix

\bibliography{apssamp}

\end{document}